\documentclass[aps,pra,twocolumn,showpacs,groupedaddress]{revtex4}  
\usepackage{graphicx}  
\usepackage{dcolumn}   
\usepackage{bm}        
\usepackage{amssymb}   
\usepackage{verbatim}  
\usepackage{epstopdf}
\usepackage{color}

\newcommand{\beq}{\begin{equation}}
\newcommand{\eeq}{\end{equation}}

\newcommand{\njp}{New. J. Phys. }

\begin{document}

\title{Antiferromagnetism in a bosonic mixture of rubidium ($^{87}$Rb) and potassium ($^{41}$K)}

\author{Uttam Shrestha\footnote{Electronic address: ushresth@uci.edu}\footnote{Present address: {\it Department of Physics and Astronomy, University of California, Irvine, California 92697-4575}}}
\affiliation{LENS, Universit\`a di Firenze, via Nello Carrara 1, 50019 Sesto Fiorentino, Firenze, Italy}

\begin{abstract}

We simulate the experimental possibility of observing the antiferromagnetic (AF) order in the bosonic mixtures of rubidium ($^{87}$Rb) and potassium ($^{41}$K) in a two-dimensional optical lattice in the presence of harmonic confinement. By tuning the interspecies interactions and the lattice heights we have found the ground states, within the mean-field approximation, that interpolate from the phase separation to the AF order. For a moderate lattice height the coexistence of the Mott and AF phase is possible for the Rb atoms while the K atoms remain in the AF-superfluid phase. This observation may provide an experimental feasibility to hitherto unobserved AF order for $^{87}$Rb - $^{41}$K mixture.
\end{abstract}

\maketitle 

The field of ultracold atoms has set a new era for its outstanding achievements for simulating many quantum phenomena related with correlated interacting particles in a strong periodic potential~\cite{Bloch}. Such systems have been successfully implemented to realize nonmagnetic phases such as Mott, superfluid and vortex states. However, realizing quantum magnetism~(QM)~\cite{Sachdev} within current experimental framework is still a challenging task  due to an extremely low value of the exchange coupling~\cite{Weld,Trotzky}, though there have been conscientious theoretical efforts and proposals in this direction~\cite{Kuklov,Duan,Altman,Hubener,Soyler,Sansone}. The possible milestone would be the experimental realization of an antiferromagentic~(AF) order in a two-component system.

Two-component boson system with tunable interspecies interactions~\cite{Thalhammer} has been a fruitful laboratory for studying many intriguing phenomena ranging from phase separation~\cite{Hall} and topological structures~\cite{Busch} to the exotic states of matter such as supersolid~\cite{Soyler} (superflow with broken translation symmetry) and counter-superfluid~\cite{Kuklov}. Current trends and publications for implementing such system in the regime where QM can be perceived are mostly theoretical, and are limited to the subspace of integer filling and to the hardcore bosons only~\cite{Soyler,Altman,Hubener}. Besides, the experimental realizations of ultracold atoms are always accompanied by the additional trapping potential that may modify the possible phase diagram.

\begin{figure}
\includegraphics[width=1.0\columnwidth]{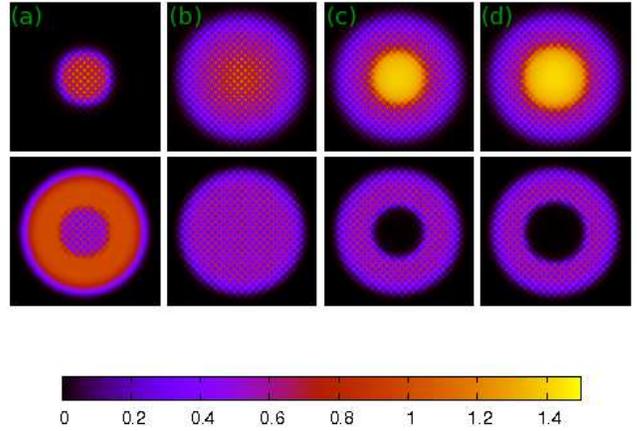}
\caption{(Color online) Density plot of $^{87}$Rb (bottom row) and $^{41}$K (top row) for $U_{\rm Rb-K}=0.28$ with $s_{\rm Rb}=14.0$ for different number of atoms $[N_{Rb},N_K]=[705,110],\,[381,441],\,[304,548],$ and $[288,610]$ for the columns (a), (b), (c), and (d). The axes represent the lattice of size $41\times 41$.}
\label {f1}
\end{figure}

In this paper we explore the ground states of a system consisting of mixture of interacting bosons in a two-dimensional~(2D) optical lattice in an external trapping potential. In particular, we focus on the recent experimental realizations of a degenerate mixture of rubidium ($^{87}$Rb) and potassium ($^{41}$K) atoms~\cite{Catani} where one can exploit the inherent asymmetry of the tunneling amplitudes and interaction energies. In general, this asymmetry is a necessary criterion for observing the AF order in the two-component system~\cite{Soyler,Altman}. For a moderate lattice potential  the heavier atoms (Rb) can be in the Mott phase while the lighter atoms (K) will remain in the superfluid phase if the interspecies interactions are weak. For stronger interspecies interactions, in place of a simple phase-separation, the lattice discreteness may trigger magnetic order such as AF phase. 

The two-component antiferromagnetism referring here is the state in which the site population alternates between two species analogous to the 2D checkerboard solid. By tuning the interspecies (Rb-K) interactions and the number of atoms of each species we have found the ground state where the density of each species alternate between adjacent sites with Rb atoms in the Mott phase while K atoms are in the superfluid phase. The domain of this checkerboard pattern extends over the size of the cloud when the number of atoms of each species are comparable (Fig.~\ref{f1}~(b)). For imbalanced mixtures the density distributions reflect the symmetry of the trapping potential with the ring like structure of Rb atoms around the densely packed K atoms as a signature of the phase separation.      

Our formulation starts with the model Hamiltonian~\cite{Jaksch},
\begin{eqnarray}
\hat H&=&\sum_{\langle i,j\rangle,\sigma}\Bigg{[}-J_\sigma(\hat{a}^\dag_{i,\sigma} \hat{a}_{j,\sigma}+\text{h.c.})+\frac{U_{\sigma,\sigma}}{2}\hat n_{i,\sigma}(\hat n_{i,\sigma}-1)\nonumber\\
&&-\mu_\sigma \hat n_{i,\sigma}+\epsilon_{i,\sigma} \hat n_{i,\sigma}\Bigg{]}+\sum_{i}
\frac{U_{\sigma,\sigma^\prime}}{2}\hat n_{i,\sigma}\hat n_{i,\sigma^\prime} \, ,
\label{Eq1}
\end{eqnarray}
where the indices $\sigma\neq\sigma^\prime$ refer to Rb or K, and $\langle i,j\rangle$ are the nearest neighbor sites. Here $\hat{a}^\dag_{i,\sigma}$ and $\hat{a}_{i,\sigma}$  are the creation and annihilation operators at the lattice site $i$ whereas $\hat n_{i,\sigma}=\hat{a}^\dag_{i,\sigma} \hat{a}_{i,\sigma}$ is the number operator for the $\sigma$ species. The parameters $J_\sigma$ and $U_{\sigma,\sigma(\sigma^\prime)}$ are respectively the tunneling amplitudes and the onsite atom-atom interactions between $\sigma$ and $\sigma(\sigma^\prime)$. The magnitude of the interspecies interactions $U_{\sigma,\sigma^\prime}$ can be varied experimentally by using Feshbach resonance. The chemical potentials $\mu_\sigma$ are adjusted to fix the number of atoms of each species. The effect of the external harmonic confinement is represented by the energy offset term $\epsilon_i$ in the Hamiltonian. In principle, Rb and K atoms may experience different trapping potentials due to their difference in masses.

For the optical lattice deep enough to validate the model~(\ref{Eq1}) the parameters can be approximately written in terms of the lattice heights and atomic scattering lengths as~\cite{Modugno}
$J_\sigma \simeq 1.4 s_{\sigma} \exp(-2.07\sqrt{s_{\sigma}})E_{r,\sigma}\, $ and
$U_{\sigma,\sigma} \simeq 5.97(a_{\sigma,\sigma}/\lambda)(s_{\sigma})^{0.88}E_{r,\sigma}\,$, where $ E_{r,\sigma}$ is the recoil energy and $s_{\sigma}$ is the lattice height of the $\sigma$ species. In the limit $J_\sigma\ll U_{\sigma,\sigma}$ and in the absence of external trapping potential ($\epsilon_i=0$), the Hamiltonian~(\ref{Eq1}) can, within the second order perturbation theory, be mapped onto effective spin-1/2 Heisenberg Hamiltonian~\cite{Kuklov}. In the simplest case when the total filling of one atom per site the effective Hamiltonian can be written as~\cite{Kuklov,Altman}
\beq
H_{\rm eff}=t_z\sum_{\langle i,j\rangle }S^z_iS^z_{j}-t_\perp\sum_{\langle i,j\rangle}(S^x_i S^x_{j}+S^y_i S^y_{j})\, ,
\eeq
where the exchange couplings scale as $t_z,\,t_\perp\sim J_\sigma^2/U_\sigma$. 
This spin model is strictly valid when the system is in deep Mott regime so that the Hilbert space is spanned by states $|n_{i,\sigma}; n_{i,\sigma^\prime}\rangle$ with $n_{i,\sigma(\sigma^\prime)}=0,1$. 
For the tunneling amplitude close to the superfluid-Mott transition the model of isospin representation breaks down. However, this gives some hints for the parameter values where one should expect the magnetic order. Since the exchange couplings scale as $t_z\sim J^2$, a large tunneling would be beneficial for providing a large exchange coupling.

In this study we use the decoupling mean-field (DMF) approximation~\cite{Sheshadri} which has been a successful model in obtaining the phase diagram of a single component system. In this approach, one can approximate the off-diagonal terms in the Hamiltonian as $
\hat{a}^\dag_{\langle i,j\rangle ,\sigma} \hat{a}_{{j,\sigma}}\approx \langle \hat{a}^\dag_i\rangle \hat{a}_{{j,\sigma}}+\hat{a}^\dag_{i,\sigma} \langle\hat{a}_{{j,\sigma}}\rangle - \langle\hat{a}^\dag_{i,\sigma}\rangle\langle \hat{a}_{{j,\sigma}}\rangle\, ,
$
where $\langle \hat{a}_{i,\sigma}\rangle\equiv\phi_{i,\sigma} $ is the so-called superfluid order parameter. Within this approximation the mean-field Hamiltonian can be written as the sum of the on-site Hamiltonians $\hat{H}_{MF}=\sum_{i} \hat{H}_{i}$, 
which can be diagonalized self-consistently using the truncated number basis. In our simulations we choose the initial order parameter $\phi_{i,\sigma}$ as real random variables in the interval $\{0,1\}$, and iterate the solution self-consistently until it converges to a desired accuracy~($10^{-8}$). In each realizations of the simulation we readjusted the chemical potentials for a given set of parameters, and the number of atoms are fixed accordingly. We have picked the converged number of atoms in the range $N_{\rm Rb}+N_{\rm K}\approx 800$ to $900$ when the solution relaxes to the ground state. In all calculations the size of the lattice is fixed to $L^2=41\times41$.  

As mentioned earlier, our study concentrates on the mixtures of $^{87}$Rb and $^{41}$K in a two-dimensional optical lattice. For the parameters in the simulations we take the masses of Rb and K atoms as $  m_{\rm Rb}=87.0 \,{\rm u}$ and $m_{\rm K}= 41.0\,{\rm u}\,$ with ${\rm u}=1.67\times10^{-27}\,{\rm kg}$, and the scattering lengths as $a_{\rm Rb}=98.0\, a_{\rm 0}$ and $ a_{\rm K}=63.0\, a_{\rm 0}\,$ with $a_{\rm 0}=0.529\times10^{-10}\,{\rm m}$. We fix the lattice height for the Rb atoms $s_{\rm Rb}=14E_{r,{\rm Rb}}$ assuming that the lattice height for the K atoms can be adjusted independently.

Figure~\ref{f1} represents an example plot of the density of K~(top row) and Rb~(bottom row) atoms for a set of four pairs of atom numbers. 
We have considered the lattice height $s_{\rm Rb}=14.0$ for the Rb 
and the corresponding  Hubbard parameters are $U_{\rm Rb}=0.30$ and $J_{\rm Rb}=0.008$, whereas for the K atoms we have taken $U_{\rm K}=0.092$, and $J_{\rm K}=0.049$ respectively. For convenience, we have taken the same trapping potentials for both the Rb and K atoms, the frequency of the Rb atoms being $\omega=2\pi\times 60 \sqrt{s_{\rm Rb}/60}$\,Hz. The frequency for the K atoms can be found by setting the condition $m_{\rm Rb}\omega_{\rm Rb}^2=m_{\rm K}\omega_{\rm K}^2$. All the energies are expressed in terms of the recoil energy of the Rb atoms. In our observations the phase separation is likely scenario if the number of atoms is large and if $U_{\rm Rb-K}>\sqrt{U_{\rm Rb}U_{\rm K}}$. For a non-lattice system the critical value of the Rb-K scattering length for the onset of the phase separation is $2\sqrt{a_{\rm Rb}m_{\rm Rb}a_{\rm K}m_{\rm K}}/(m_{\rm Rb}+m_{\rm K})$~\cite{Riboli}. It should also be noted that the critical value for the superfluid-Mott insulator transition in a homogeneous system with unit filling is $(U_\sigma/zJ_\sigma)_c=5.8$~\cite{Fisher}. If we neglect the effect of the trapping potentials the Rb atoms remain in the Mott insulating (if the integer filling prevails) while the K atoms remain in the superfluid phase. 
 
The density distribution in Fig.~\ref{f1} form a ring of Rb atoms around the dense K atoms at the center of the trap. It is expected since Rb atoms are interacting stronger than K atoms and are pushed outward from the center of the trap. For small number of K atoms a small patch of AF order is formed near the center of the cloud surrounded by an extended Mott plateau of Rb atoms. As the number of K atoms increases the AF order extends over the whole lattice, and the correlator $\langle a_\sigma ^\dag a_\sigma\prime \rangle$ shows minimum when $N_{\rm Rb}\approx N_{\rm K}$. For large number of K atoms the overlap region shrinks in the form of ring near the periphery while the central region is void of Rb atoms. In both extreme cases we have signs of phase separation along with the existence of the magnetic order.   

\begin{figure}
\includegraphics[width=1.0\columnwidth]{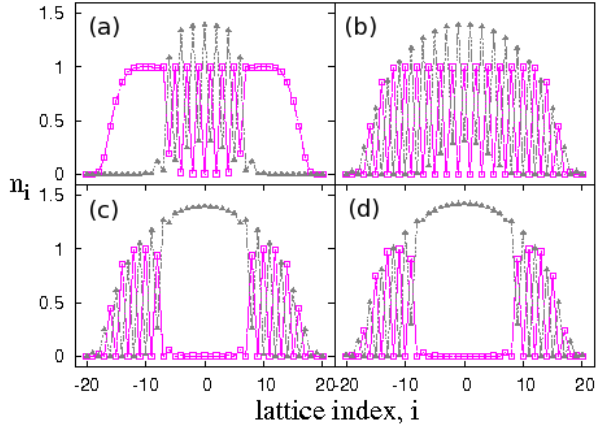}
\caption{(Color online) Atom distribution for the Rb (square) and K (triangle) atoms along the line $x=0$. The data is extracted from the Fig.~\ref{f1}.}
\label {f2}
\end{figure}

In Fig.~\ref{f2} we plot the cut of the density profile along the line $x=0$ for the same parameters as in Fig.~\ref{f1}. The Mott plateau of Rb and superfluid profile of the K atoms exist along with the AF order in the overlapped region.

In order to detect the hidden order in the lattice experimentally a number of proposals have been put forward. Two prominent techniques are the noise correlation~\cite{Noise,Folling,Fallani} and the lattice modulation spectroscopy~\cite{Kollath}. Here we emphasize the noise correlation by analyzing the images of the density distribution during the ballastic expansion of the cloud in the far field regime~\cite{Folling,Fallani}. 

We define the normalized correlation function in the momentum space
\beq
C({\vec k})= \frac{\int d{\vec q}\, \langle \hat n({\vec q})\hat n({\vec q+\vec k})\rangle}{\int d{\vec q}\, \langle \hat n({\vec q})\rangle \langle \hat n({\vec q+\vec k})\rangle}\, .
\eeq 

If the state of the system is known, the expectation values that appeared in the integrals can be found easily. In particular, if the state is defined by the product of the number state 
$|\phi_0\rangle={\Pi}_i |n_i\rangle\, ,$
which is a valid assumption for Rb atoms, the correlation function turns out to be
\beq
C_{\rm Rb}({\vec k})=\frac{\sum_i n_{i,\rm Rb} ^2}{N_{\rm Rb} ^2}-\frac{1}{N_{\rm Rb} ^2}+\frac{n_{\vec k, {\rm Rb}} ^2}{N_{\rm Rb}^2} \, .
\eeq
Here we have neglected the term containing the delta function.

\begin{figure}
\includegraphics[width=1.0\columnwidth]{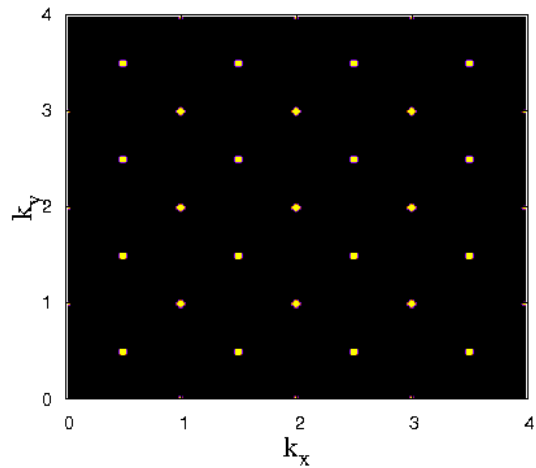}
\caption{Noise correlation function $C_{\rm Rb}({\bf k})$ for Rb for the parameters of Fig.~\ref{f1} (b). Mott peaks of the original lattice are spanned by the vector $\vec k=l(n\hat x,m\hat y)$ whereas the additional peaks are at the position $\vec k=l/\sqrt{2}(-n\hat x,m\hat y)$ where $n$ and $m$ are integers and $l=2\pi/|d|$, $d$ being the lattice constant.}
\label {f3}
\end{figure}

In Fig.~\ref{f3} we show the density plot of the correlation function $C_{\rm Rb}({\vec k})$ in the $k_x-k_y$ plane corresponding to the parameters of Fig.~\ref{f1} (b). In addition to the Mott peaks for integer values of ($k_x,k_y$) the correlation function develops additional peaks. It should be noted that for the checker-board solid the size of the unit cell in the real space gets doubled and rotated by $\pi/2$ with respect to the original lattice. It can be easily verified that the length of the unit vector in the reciprocal lattice is $1/\sqrt 2 $ times that of the original lattice. Therefore, the secondary peaks in the correlation function do not appear at the position half integral multiple of  $k_x$ or $k_y$, but at the center of the unit cell spanned by the original reciprocal lattice vector. For this particular set of parameters in Fig.~\ref{f3} the ratio of the amplitudes of the secondary ($\vec k=1/2$) to the primary ($\vec k=0$) peaks in the correlation function is $0.45$.    

Another possibility to observe the AF order in the two component system is to create an excitation by means of lattice shake or by modulating the intensity of the lattice height periodically. The AF order can be revealed with an additional peak of $U_{\rm Rb-K}$ in the lowest excitation spectrum~\cite{Kollath}. This is because the lattice modulation transfer a Rb atom to the nearest occupied lattice site and the gain in energy will be $U_{\rm Rb-K}$ if the nearest site is occupied by the K atom. This approach can be suitable for a strong lattice confinement with both components forming Mott insulating states. But for the parameters we are dealing in this paper the number of K atoms in a given site is not well defined due to the atom number fluctuations and the lattice modulation may give a broad peak in the excitation spectrum. 

\begin{figure}
\includegraphics[width=1.0\columnwidth]{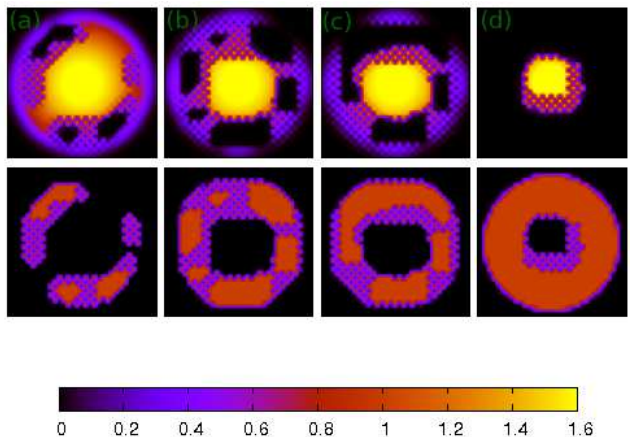}
\caption{(Color online) Density plot of $^{87}$Rb (bottom row) and $^{41}$K (top row) for $U_{\rm Rb-K}=0.45$ with $s_{\rm Rb}=20.0$. The number of atoms are $[N_{Rb},N_K]=[188,640],\, [354,458], \,[580,200],$ and $[665,100]$ for columns (a), (b), (c), and (d). The axes represent the lattice of size $L^2=41\times41$.}
\label {f4}
\end{figure}

In order to refine our analysis we have also varied the trapping potentials for the Rb and K separately. In this case we have not observed a clear signature of the AF order but the phase separation only. This may be due to the fact that the difference in local chemical potentials between Rb and K atoms greatly exceed the tiny exchange interaction needed to stabilize the AF order. In the experimental situation the equal trapping potential may not be a problem as the frequencies can be tuned by setting the condition $m_{\rm Rb}\omega_{\rm Rb}^2=m_{\rm K}\omega_{\rm K}^2$. 

To be more concrete with our speculation that the large hopping amplitude could be a regime conducive for the AF order we consider a deeper optical lattice with $s_{\rm Rb}=20$ for the Rb. In Fig.~\ref{f4} we show the density plot for the typical atom numbers of each species when the interspecies interactions is $U_{\rm Rb-K}=0.45$. The simulation parameters are $U_{\rm Rb}=0.41$, $U_{\rm K}=0.13$, $J_{\rm Rb}=0.0025$, and $J_{\rm K}=0.026$. The trapping frequency for the Rb atoms is $\omega=2\pi\times 60 \sqrt{s_{\rm Rb}/60}$\,Hz while the frequency of the K atoms is found by the equal potential conditions. The Mott plateau of the Rb atoms and the superfluid profile of the K atoms still persist but the AF order washes away. There are few patches of the AF order in the region where two species overlap.

From the theoretical point of view when both species are in deep Mott regime there could be an AF order but the corresponding exchange couplings are extremely low.  In the deep Mott regime atom tunneling is exponentially suppressed, and a tiny perturbation such as thermal and quantum fluctuations are sufficient to destroy the fragile AF order. Our observation is also consistent with the previous phase diagram~\cite{Soyler} of two-component bosons on a square lattice at half-integer filling of each species. For the parameters of Fig.~\ref{f1} we have $2zJ_{\rm Rb}/U_{\rm Rb-K}=0.22$ and $2zJ_{\rm K}/U_{\rm Rb-K}=1.4$, which lies in the 2CB-phase of Fig.~1 in Ref.~\cite{Soyler}.

In~\cite{Catani} it is reported that the minor fraction of the K atoms reduces the visibility of the interference pattern of the Rb atoms, we speculate that this impurity-induced loss of coherence~\cite{Catani,Gadway} may be accounted for the presence of the AF order, and the detail experimental findings in the regime of comparable atom number of each species would reveal more information. 

Unlike in a single component system where the chemical potential fixes the total number of atoms, the two-component system allows number fluctuations for the same set of chemical potentials when the interspecies interactions are switched on. 
The asymmetry observed in the density distribution in Fig.~\ref{f4} may be due to the degeneracy present in the ground state since a set of chemical potentials may give the different set of atom numbers. It is interesting to compare the quantum results shown in Fig.~\ref{f4} with the experimental demonstration of simultaneous existence of superfluidity and magnetism in the spinor Bose-Einstein condensate~\cite{Sadler}.  

In summary, we have simulated a typical scenario of a degenerate bosonic mixtures of the Rb and K atoms in a 2D optical lattice in the presence of external trapping potential. By tuning the interspecies interactions we find that the AF order is possible if the Rb atoms are in the Mott phase while K atoms are in the superfluid phase. This ordered phase can be reflected in the density distribution as well as in the noise-correlation signal. In our view there has not been any study (at least to our knowledge) on the AF order in the two-component system when one species remain in the superfluid phase while the other in the Mott phase, this observation may provide the novel regime for studying quantum magnetism in ultracold system.

We would like to thank F. Minardi,  J. Catani and M. Modugno for fruitful comments and suggestions. This work was supported by the EU Contract EU STREP NAMEQUAM.

{}


\begin{thebibliography}{99}

\bibitem{Bloch}
I.~Bloch {\it et al.}, \rmp {\bf 80}, 885 (2008);
M.~Lewenstein {\it et al.}, Adv. Phys. {\bf 56}, 243 (2007).
\bibitem{Sachdev}
S.~Sachdev, Nat. Phys. 4, 173 (2008).
\bibitem{Weld}
D.M.~Weld {\it et al.},  \prl {\bf 103}, 245301 (2009).
\bibitem{Trotzky}
S.~Trotzky {\it et al.}, Science {\bf 319}, 295 (2008).
\bibitem{Kuklov}
A.B.~Kuklov and B.V.~Svistunov,  \prl {\bf 90}, 100401 (2003).
\bibitem{Duan}
L.M.~Duan {\it et al.},  \prl {\bf 91}, 090402 (2003). 
\bibitem{Altman}
E.~Altman {\it et al.},  \njp {\bf 5}, 254109 (2003).
\bibitem{Hubener}
A.~Hubener {\it et al.},  \prb {\bf 80}, 254109 (2009).
\bibitem{Soyler}
S.G.~S$\ddot {\rm o}$yler {\it et al.},  \njp {\bf 11}, 073036 (2009).
\bibitem{Sansone}
B.~Capogrosso-Sansone {\it et al.},  \pra {\bf 81}, 053622 (2010).
\bibitem{Thalhammer}
G.~Thalhammer {\it et al.},  \prl {\bf 100}, 210402 (2008).
\bibitem{Hall}
D.S.~Hall {\it et al.},  \prl {\bf 81}, 1539 (1998);
E.~Timmermans, Phys. Rev. Lett. {\bf 81}, 5718 (1998);
S.B.~Papp {\it et al.}, \prl {\bf 101}, 040402 (2008).
\bibitem{Busch}
Th.~Busch and J.R.~Anglin, \prl {\bf 87}, 010401 (2001);
J.~Ruostekoski and J.R.~Anglin, Phys. Rev. Lett. {\bf 86}, 3934 (2001);
U.~Shrestha {\it et al.}, \prl {\bf 103}, 190401 (2009).
\bibitem{Catani}
J.~Catani {\it et al.}, \pra {\bf 77}, 011603(R) (2008).
\bibitem{Jaksch}
D. Jaksch {\it et al.}, Phys. Rev. Lett. {\bf 81}, 3108 (1998). 
\bibitem{Modugno}
F.~Gerbier {\it et al.}, \pra {\bf 72}, 053606 (2005); M.~Modugno, \njp~{\bf 11}, 033023 (2009).
\bibitem{Sheshadri}
K.~Sheshadri {\it et al.}, Europhys. Lett. {\bf 22}, 257 (1993); 
K.~Sheshadri {\it et al.}, \prl {\bf 75}, 4075 (1995).
\bibitem{Riboli}
F. Riboli and M. Modugno, \pra~{\bf 65}, 063614 (2002).
\bibitem{Fisher}
M.P.A.~Fisher {\it et al.}, \prb~{\bf 40}, 546 (1989).
\bibitem{Noise}
E.~Altman {\it et al.}, \pra. {\bf 70}, 013603 (2004).
\bibitem{Folling}
S.~F$\ddot {\rm o}$lling {\it et al.}, Nature (London) {\bf 434}, 491 (2005).
\bibitem{Fallani}
L.~Fallani {\it et al.}, \prl {\bf 98}, 130404 (2007).
\bibitem{Kollath}
C.~Kollath {\it et al.}, \prl {\bf 97}, 050402 (2006); T.~St$\ddot {\rm o}$ferle {\it et al.}, \prl {\bf 92}, 130403 (2004); 
\bibitem{Gadway}
B.~Gadway {\it et al.},  \prl {\bf 105}, 045303 (2010).
\bibitem{Sadler}
L.E.~Sadler {\it et al.}, Nature (London) {\bf 443}, 312 (2006). 
\end{thebibliography}
\end{document}